\documentclass[a4paper]{spie}  %>>> use this instead for A4 paper
\usepackage[]{graphicx}

\title{Focusing of gamma-rays with  Laue lenses: first results} 

\author{F. Frontera\supit{a,c}, G. Loffredo\supit{a}, A. Pisa\supit{a}, 
F. Nobili\supit{a},V. Carassiti\supit{b}, F. Evangelisti\supit{b}, L. Landi\supit{a}, 
S. Squerzanti\supit{b}, E. Caroli\supit{c}, J.B. Stephen\supit{c},  
K.H. Andersen\supit{d}, P. Courtois\supit{d}, N. Auricchio\supit{a}, 
L. Milani\supit{a}, B.~Negri \supit{e}
\skiplinehalf
\supit{a}University of Ferrara, Physics Department, Via Saragat 1, 44100
Ferrara, Italy; \\
\supit{b}Istituto Nazionale Fisica Nucleare, Sezione di Ferrara, Via Saragat 1,
44100 Ferrara, Italy; \\
\supit{c}INAF, IASF Bologna, Via Gobetti 101, 40129 Bologna, Italy\\
\supit{d}Institute Laue--Langevin, 6 Rue Jules Horowitz, 38042 Grenoble, France\\
\supit{e}Agenzia Spaziale Italiana, Viale Liegi, 26, 00198 Roma, Italy\\
}

\authorinfo{Further author information: (Send correspondence 
to F.F.)\\ F.F: E-mail: frontera@fe.infn.it, Telephone: +39 0532 974 254}

\begin{document} 
  \maketitle 

%%%%%%%%%%%%%%%%%%%%%%%%%%%%%%%%%%%%%%%%%%%%%%%%%%%%%%%%%%%%% 
\begin{abstract}
We report on the first results obtained from our development 
project of focusing gamma-rays ($>$60 keV) by using Laue lenses. The first 
lens prototype model has been assembled and tested. We describe the
technique adopted and the lens focusing capabilities at about 100 keV.

\end{abstract}


%%%%%%%%%%%%%%%%%%%%%%%%%%%%%%%%%%%%%%%%%%%%%%%%%%%%%%%%%%%%%
\section{INTRODUCTION}
\label{s:intro}  

Experimental hard X--/gamma--ray astronomy is moving from direct 
sky-viewing telescopes to focusing telescopes.
With the advent of focusing telescopes in this energy range, a 
big leap in sensitivity is expected, by a factor of 100-1000 with respect
to the best non-focusing instruments of the current generation, either 
using coded masks
(e.g., INTEGRAL/IBIS, Ref.~\citenum{Ubertini03}) or not (e.g., BeppoSAX/PDS, 
Ref.~\citenum{Frontera97}). 
A significant increase in angular resolution is also expected
(from $\sim 10$ arcmin of the mask telescopes like INTEGRAL IBIS to 
less than 1 arcmin).

The next generation of hard X--($<$100keV) and gamma--ray ($>$ 100 keV) 
focusing telescopes will make use of the Bragg diffraction technique, 
in the former energy range, from 
multilayer coatings (ML) in a reflection configuration (supermirrors), while
in the latter, from mosaic--like crystals in a transmission configuration 
(Laue lenses). 

The expected astrophysical issues that are expected to be solved with the advent
of these telescopes are many and of fundamental importance.
For what concerns the higher energy band ($>$100 keV), a summary of the 
main science goals are discussed in the context of a mission 
proposal, {\em Gamma Ray Imager} (GRI), submitted to ESA in response to 
the first AO of the 'Cosmic Vision 2015--2025' plan \cite{Knodlseder07}. For the 
astrophysical importance of the $>$100 keV band, see also
Refs.~\citenum{Frontera05a,Frontera06,Knodlseder06,Knodlseder07}. 
The GRI proposal was not admitted by ESA for further consideration for a 
launch in 2017
(the first flight opportunity in the new ESA plan) mainly due to readiness problems
of the Laue lenses. 

Here we report on the status of 
our project HAXTEL (= HArd X-ray TELescope) devoted to  developing the technology 
for building broad energy passband (70/100--600 keV) Laue lenses, highlighting 
test results of the first lens prototype.

%%%%%%%%%%%%%%%%%%%%%%%%%%%%%%%%%%%%%%%%%%%%%%%%%%%%%%%%%%%%%

\section{Summary of the lens design study results}
The results of the Laue lens development activity over the last few years
have been reported and discussed 
(see Refs.~\citenum{Frontera06,Frontera07} and references therein).

In short, by means of theoretical calculations and the development of a 
Monte Carlo code, we first established the best geometrical configuration 
of a Laue lens for space astronomy, the best crystal 
materials, and the crystal mosaic properties that give the best compromise between 
reflection efficiency and focusing capabilities. This theoretical study 
was followed by reflectivity measurements of mosaic crystal samples of 
Cu(111) \cite{Pellicciotta06}, given that this material showed 
good expected performance for our goals and could be developed 
with the desired properties (e.g., Courtois et al.\cite{Courtois04}). 

Given that a Laue lens focusing telescope for gamma--ray astronomy is required
to have a spherical--like shape in addition to a large size (meter scale), 
while the available mosaic crystal sizes are generally small (cm scale), 
the only viable solution is to build a lens made of many crystal tiles. This means 
that, in order to correctly focus the photons reflected from all 
crystal tiles toward the same point, the direction of the 
vector perpendicular to the mean lattice plane of each crystal tile has to intersect 
the lens axis, while its inclination with respect to the focal plane has 
to be equal to the Bragg angle $\theta_B$ (see Fig.~1 in 
Ref.~\citenum{Pellicciotta06}).
The angle $\theta_B$ also establishes the energy centroid of the reflected photons,
and it depends on the distance $r$ of the tile center from the lens axis 
and on the focal length $f$. For correct focusing,
$\theta_B = 1/2 \arctan{(r/ f)}$.
Once the crystal material is established, the Bragg angle increases with $r/f$, 
while the energy of the focused photons decreases with $r/f$.
More generally, once the focal length is established, the outer  and inner
lens radii, $r_{max}$ and $r_{min}$, depend on the
nominal energy passband of the lens ($E_{min}$, $E_{max}$) and on the crystal 
lattice spacing: higher $d_{hkl}$ implies lower radii \cite{Frontera06}. For a
given crystal material, outer lens radius and focal length, the minimum energy 
that can be focused is established.

For a given inner and outer radius, the lens passband  can
be established by  the combination of the different crystal materials used.
In addition to Copper, other candidate materials, which exhibit high 
reflectivity and for which the mosaic technology has been developed, 
are becoming available for Laue lenses, 
such as mosaic Ge, Si$_{1-x}$Ge$_x$, with a 
composition--gradient \cite{Abrosimov05}, or mosaic Gold \cite{Barriere08}.
 
Mosaic spread and crystal thickness are the most crucial 
parameters for the optimization of the lens performance. 
The crystal thickness issue was investigated by us in relation to the
lens weight (e.g., Ref.~\citenum{Pisa05b}). Concerning the mosaic spread, a 
higher spread gives a larger effective area, but also produces a larger defocusing 
of the reflected photons in the focal plane. We introduced a parameter that 
characterizes a Laue lens and its sensitivity: the focusing factor $G$
\begin{equation}
G = f_{ph} \frac{A_{eff}}{A_d}
\end{equation}
 in which $A_{eff}$ is the effective area of the lens and $A_d$ is the
area of the focal spot which contains a fraction $f_{ph}$ of photons 
reflected by the lens. The higher $G$ is, the higher is the lens sensitivity 
(minimum detectable source flux $F_{min} \propto 1/G$).

The spatial distribution of the mosaic crystal tiles influences the shape of the
effective area vs. energy. For low focal lengths ($< \sim10$~m), the best crystal 
tile disposition is an Archimedes' spiral \cite{Pisa04}, that provides a smooth 
behavior of the lens effective area $A_{eff}$ with energy. For high focal lengths, 
a ring crystal disposition is equally good.

The accuracy of the crystal tile positioning in the lens
depends not only on the mosaic spread but also 
on the focal length. Higher focal lengths require a higher positioning accuracy. 
At the current stage of development, an accurate positioning of the crystals 
in the lens is one of the major problems to be faced.

\begin{figure}
\begin{center}
\includegraphics[angle=0, width=0.4\textwidth]{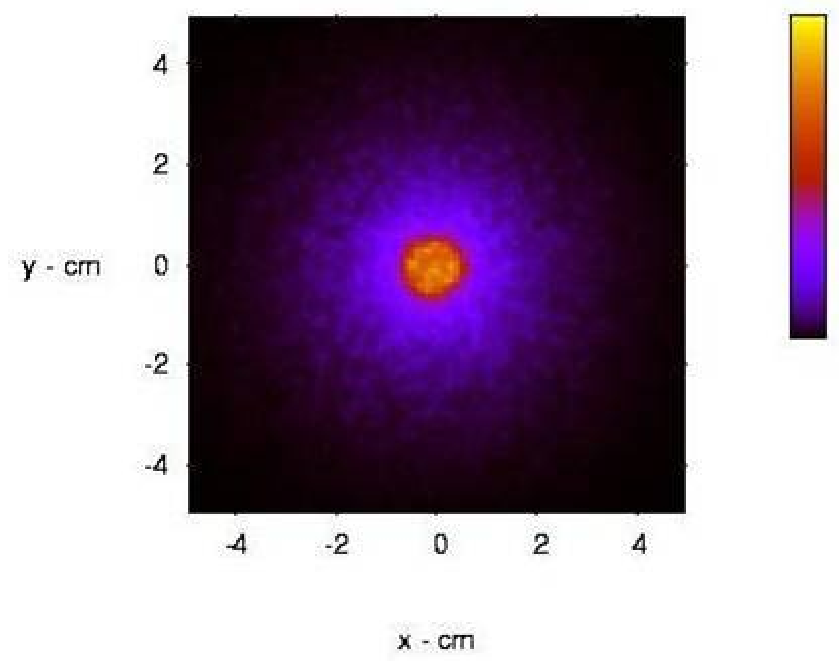}
\includegraphics[angle=0, width=0.4\textwidth]{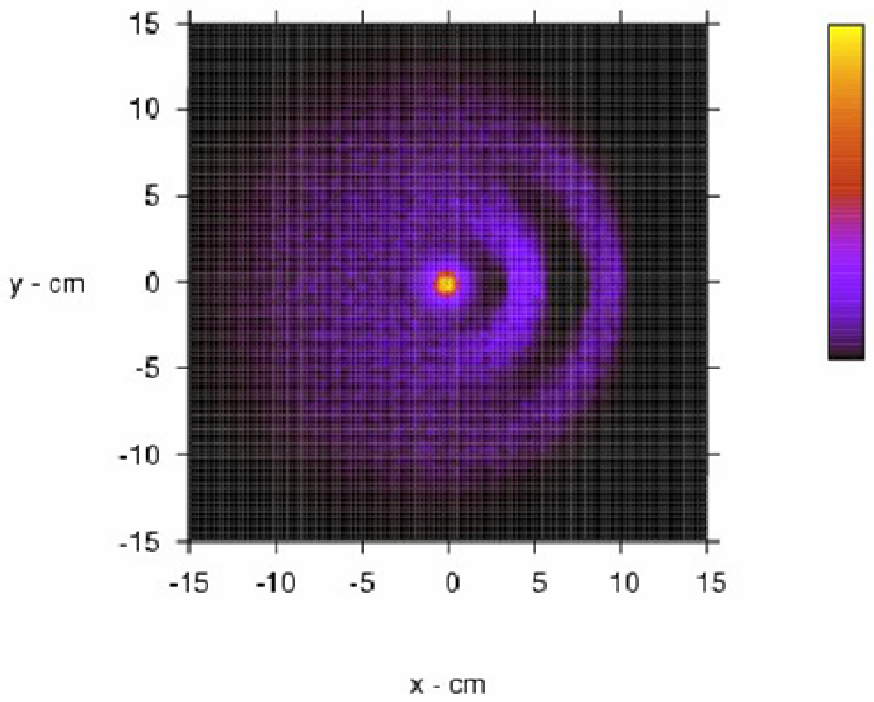}
\end{center}
%\vspace{-0.5cm}
\caption{Expected response function PSF of a Laue lens {\em Left panel:} 
on-axis PSF.  {\em Right panel:} off--axis PSF, in which three sources are
simulated at 0, 2 and 4 arcmin off--axis. See text for details.}
\label{f:psf}
\end{figure}

From Monte Carlo (MC) studies, we have derived the  
properties of various lens configurations, such as their Point Spread Functions (PSF), 
for both on--axis and off--axis incident photons.
We show in Fig.~\ref{f:psf} the expected  PSF of a Cu (111) gamma--ray Laue lens 
with a focal length of 40 m and a mosaic spread of 1 arcmin, in which
a perfect positioning of the composing crystals is assumed. 
The  expected  angular resolution of this lens is of the order of or even better 
than 1 arcmin.

%
% Figure 1 
%

\section{Prototype Model Development}

The first Laue lens Prototype Model (PM) has been developed. Unlike the Laue lens 
described in Ref.~\citenum{Vonballmoos04}, the PM developed is made of crystal tiles 
rigidly fixed to the lens frame, without any  mechanism for 
adjustment of their orientation in the lens, thus their correct positioning 
in the lens is performed during the lens assembly. 
The goal of the first PM is to test the lens assembling technique adopted.

\subsection{Lens assembly technique}

Details of the lens assembling steps have already been reported \cite{Frontera07}. 
The adopted lens assembling technique is based on the use of a counter-mask 
provided with holes, two for each crystal tile. Each tile is placed
on the countermask by means of two pins, rigidly glued to the tile, that
are inserted in the countermask holes. The pin direction and the axis
of the average lattice plane of each crystal tile have to be exactly
orthogonal. The hole direction constrains the energy
of the photons diffracted by the tile, while the relative position of
the each two holes in the countermask
establishes the azimuthal orientation of the mean crystal lattice plane,
that has to be othogonal to the lens axis.
Depending on the countermask shape and, mainly on the direction of the hole 
axes, the desired geometry of lens can be obtained. 
In the case of a lens for space astronomy, the hole axes have to be all
directed toward the center of curvature of the lens. 

In the case of the developed PM, the hole axes are set parallel to the
of the lens axis. This choice has been made to illuminate the entire lens with the
available test gamma--ray beam, which is isotropic, being generated by an X--ray
tube, and also highly divergent, as the source is only  a small distance 
away ($d \sim 6$~m) from the lens.

Once the average direction of the chosen lattice planes (111) of
each Cu mosaic crystal tile has been determined, the two pins, inserted in a pin
holder, are glued to the lens. The pin holder direction is preliminarly
made parallel to the gamma--ray beam axis and thus to the direction of the 
average crystalline plane chosen (in our case, the (111) planes). 

Once all the crystal tiles are placed on the counter-mask, a
frame is glued to the entire set of the crystals. Then the lens frame, along
with the crystals, is separated 
from the counter-mask and from the pins by means of a chemical etching that 
dissolves the aluminum caps that cover the pin bases glued to each crystal.

\begin{figure}
\begin{center}
\includegraphics[angle=0, width=0.4\textwidth]{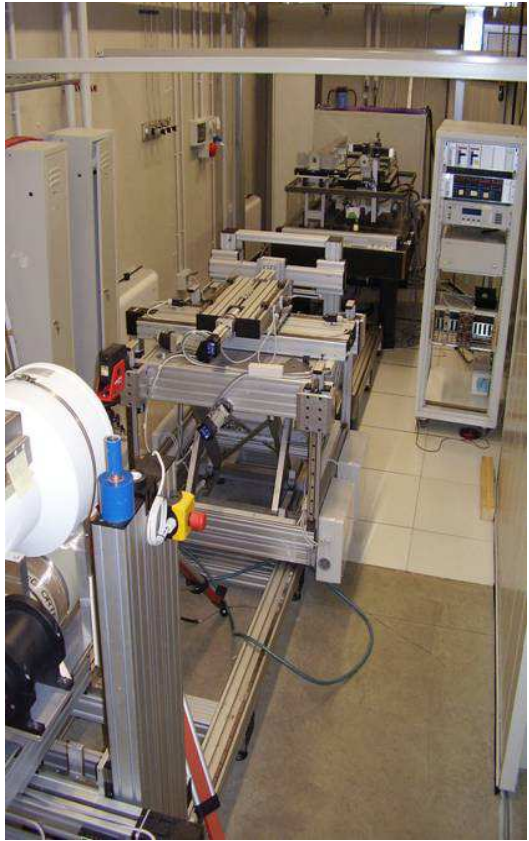}
\end{center}
%\vspace{-0.5cm}
\caption{A view of the apparatus for assembling  the
lens PM. The apparatus is located in the LARIX lab of the University
of Ferrara. }
\label{f:facility}
\end{figure}

The average direction of the chosen crystalline planes and of the pin axes
are determined by means of an X--ray beam developed for this project and located
in the LARIX (LArge Italian X--ray facility) lab of the University of Ferrara
For a LARIX description see Ref.~\citenum{Loffredo05}). 
A view of the experimental apparatus used to assemble and test the  lens PM
is shown in Fig.~\ref{f:facility}. 
%
% Figure 2
%
As discussed in Ref.~\citenum{Frontera07}, the developed apparatus includes 
an X--ray generator tube with a fine focus of 0.4 mm radius, a  maximum 
voltage of 150~kV and a maximum power of $\sim 200$~W.

The photons coming out from the gamma--ray source box
are first collimated, with the beam axis made horizontal and directed toward the
centre of a large (30 cm diameter) X--ray imaging detector whose
position sensitivity is 300~$\mu$m.
The collimator aperture can be remotely adjusted in two orthogonal directions
in order to have the desired X--ray beam size \cite{Frontera07}.   
In addition to the X--ray imaging detector, a cooled HPGe detector and 
a position sensitive (2 mm) scintillator detector are available.

For each mosaic crystal tile, the direction of the average lattice plane 
was determined with an accuracy better than 10 arcsec \cite{Frontera07},
while the alignment of the pin axis with the beam direction
was performed with an estimated accuracy of 1 arcminute. 

\section{Developed PM}

\begin{figure}
\begin{center}
\includegraphics[angle=0, width=0.4\textwidth]{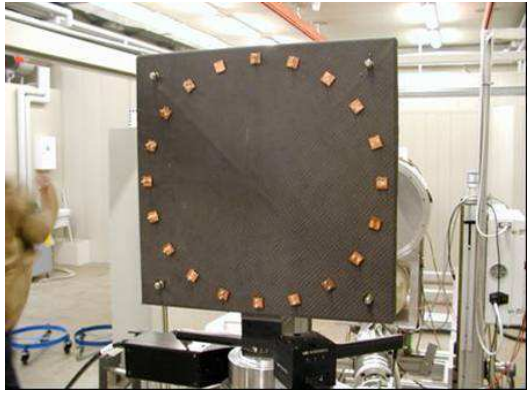}
\end{center}
%\vspace{-0.5cm}
\caption{Laue lens prototype model (PM) mounted on its holder to
being tested.}
\label{f:PM}
\end{figure}

The mounted PM is shown in Fig.~\ref{f:PM}. The PM is composed of a 
ring of 20 mosaic crystal tiles of Cu(111). The ring diameter is 36 cm. The 
mosaic spread  of the used crystals ranges from $\sim 2.5$ and $\sim 3.5$ arcmin.
The tile cross--section is 15$\times$15 mm$^2$ while its thickness is 2 mm. 
The lens frame is made of carbon fiber of 1 mm thickness.

%
% Figure 3
%

\begin{figure}
\begin{center}
\includegraphics[angle=0, width=0.4\textwidth]{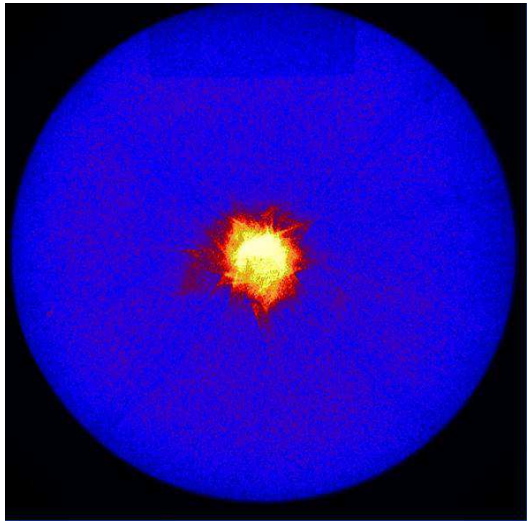}
\end{center}
%\vspace{-0.5cm}
\caption{PSF obtained from the first light of the lens PM. 
The integration time is 2 s}
\label{f:first_light}
\end{figure}

The PM was widely tested using the polychromatic X--ray beam described above.
To this end the lens  was positioned on the same holder used to determine the 
direction of the crystalline planes of each crystal tile. The detectors, mounted
on a frame whose distance from the lens can be adjusted, were
positioned at the same distance of the X--ray source from the lens ($\sim6$~m),
where the focal plane is expected. 
In order to avoid that direct radiation could arrive onto the focal plane detectors,
a lead shield 3 mm thick covered the entire inner lens frame. 
The background radiation was measured when the X--ray tube was switched on, with 
the X--ray beam stopped before arriving to the lens plane.
Figure~\ref{f:first_light} shows the first light of the lens when the 
X--ray polychromatic beam irradiates all the lens crystals. 

%
% Figure 4
%

\begin{figure}
\begin{center}
\includegraphics[angle=0, width=0.4\textwidth]{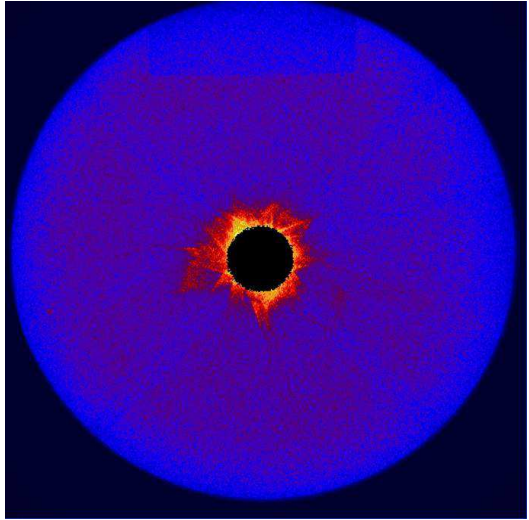}
\end{center}
%\vspace{-0.5cm}
\caption{Difference between the PSF shown in Fig.~\ref{f:first_light}
and that obtained with a Monta Carlo code by assuming a perfect positioning 
of the crystal tiles in the lens.}
\label{f:difference_image}
\end{figure}

In order to compare the measured PSF with that expected, we have developed 
a Monte Carlo code that simulates the same lens, in which the mosaic crystal 
tiles are perfectly positioned. The PSF of this lens has been derived for
the same diverging X--ray beam. 
The difference between the measured PSF of Fig.~\ref{f:first_light}
and the simulated one is shown in Fig.~\ref{f:difference_image}.
As can be seen, only the center part of the measured image (the black region) 
is  subtracted by the simulated image. The corona still visible
in the difference image is the result of the cumulative error made in the 
crystal tile positioning.
%
% Figure 5
%

\begin{figure}
\begin{center}
\includegraphics[angle=0, width=0.4\textwidth]{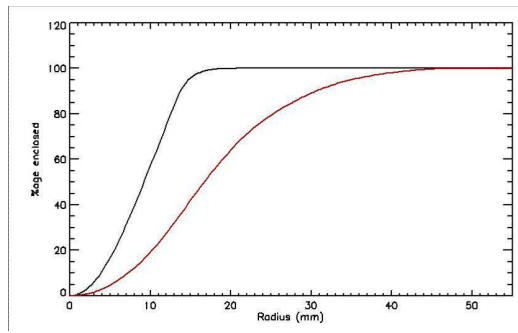}
\end{center}
%\vspace{-0.5cm}
\caption{Cumulative distribution of the focused photons with the radial 
distance from the focal point. {\em Black line:} expected distribution in the case
 of a perfect mounting of the crystal tiles in the lens. 
{\em Red line:} measured distribution.}
\label{f:spread}
\end{figure}

The disagreement between the measured and the expected PSF has also been 
obtained by comparing their radial profiles. 
The result is shown in Fig.~\ref{f:spread}. As it can be seen, the
radius at which the expected
cumulative distribution of the focused photons reach the saturation, corresponds
to the radius at which  60\% of the focused photons are collected. 

%
% Figure 6
%

\begin{figure}
\begin{center}
\includegraphics[angle=0, width=0.4\textwidth]{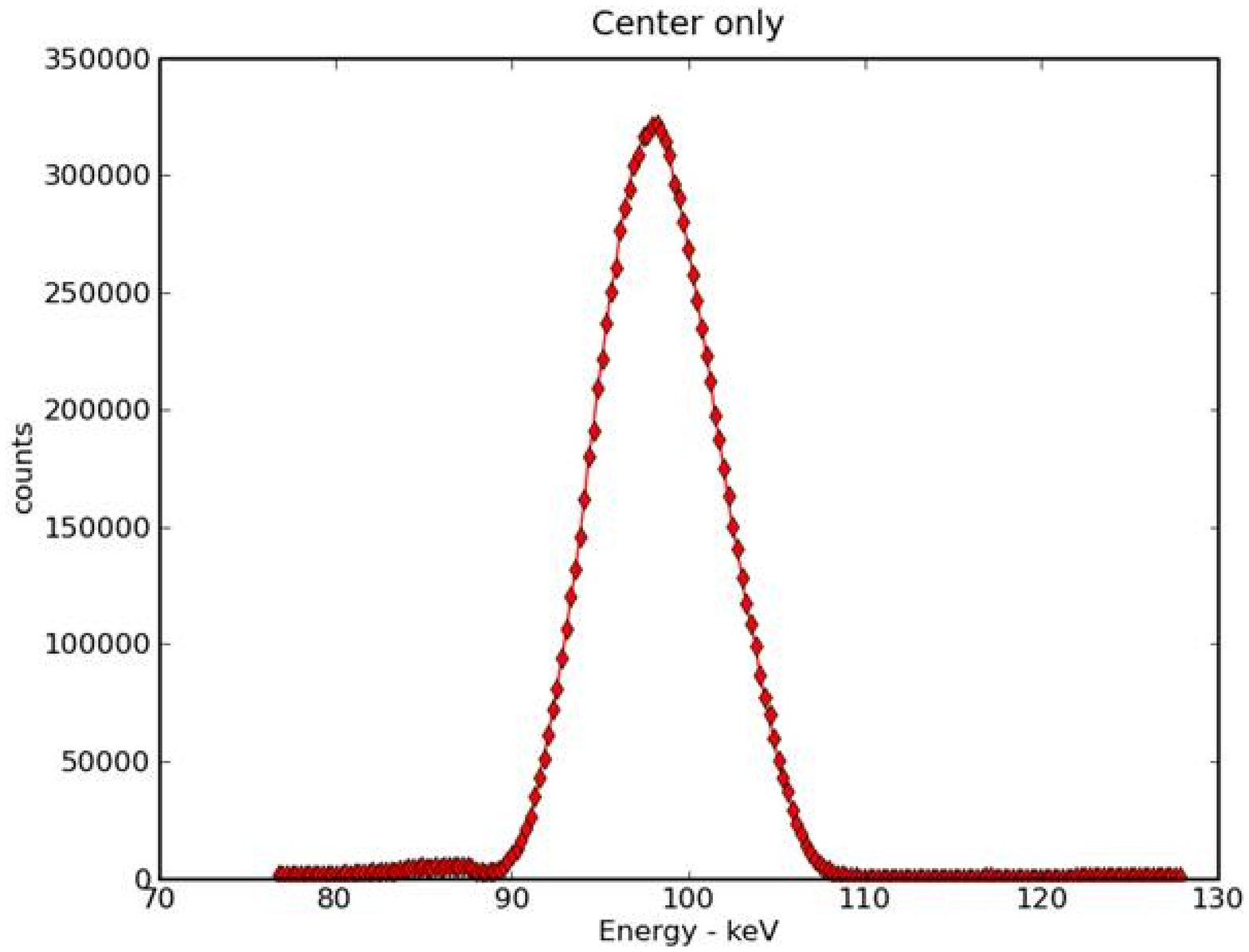}
\includegraphics[angle=0, width=0.4\textwidth]{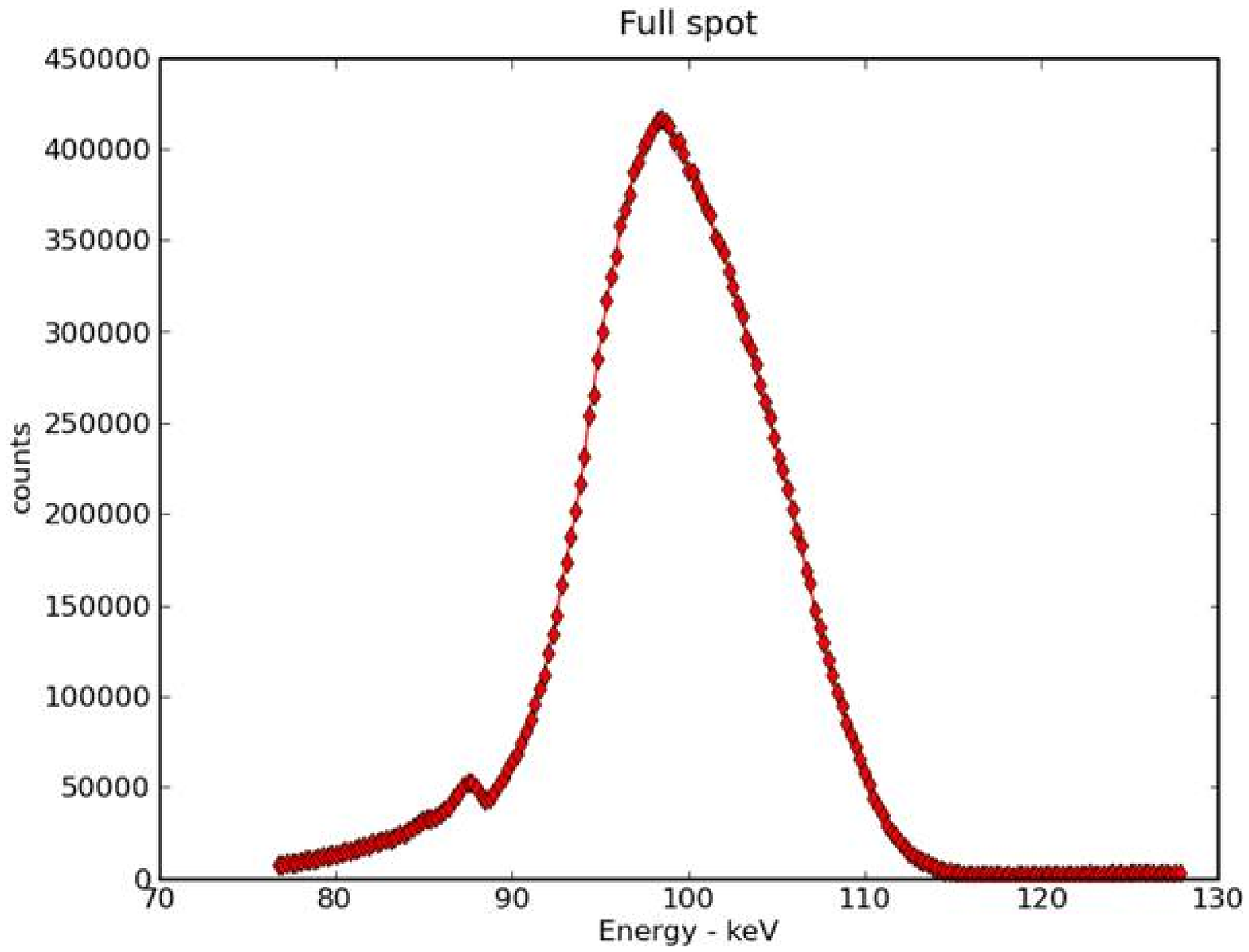}
\end{center}
%\vspace{-0.5cm}
\caption{{\em Left panel:} photon spectrum of the region (see 
Fig.~\ref{f:difference_image}) in which all the reflected photons 
are expected to be found in the case of a perfect mounting of the 
crystal tiles in the lens. 
{\em Right panel:} spectrum of all reflected photons.}
\label{f:spectrum}
\end{figure}

 Also the spectrum of the developed PM has been measured. Given the distance 
of the diverging X--ray source and its distance from the lens, from the radius
of the lens ring, the spectrum of the reflected  photons is expected to show
a peak around 100 keV. This indeed is what is found, as can be seen from 
Fig.~\ref{f:spectrum}, where we compare the measured spectrum of the central
region (see black region in Fig.~\ref{f:difference_image} with the spectrum
of all reflected photons (see Fig.~\ref{f:first_light}). As can also be seen 
from this figure, the centroid of the spectrum of the central region
achieves an intensity level 0.8 times that of the peak spectrum of all
reflected photons.

%
% Figure 7
%

Also the effect of the temperature on the focusing properties of the lens
have been investigated, finding a significant defocusing when the environment
temperature has been decreased by 6 degrees with respect to the mean value (25.5 
$^\circ$C) at which the lens assembly procedure was performed.

\section{Analysis of the results and work in progress}

By shielding all the lens cystals but one, we have investigated 
the contribution of each lens crystal to
the lens PSF and to the reflected photon spectrum. From these results, we have 
derived the positioning error of each crystal in the lens.
In order to establish the true contribution of each assembling step to the 
cumulative error, we have performed further 
tests. In this way we have estimated the precision achieved in the 
single assembling steps.

From this analysis we have identified  the likely origin of the positioning
errors and established actions to be taken in order
to remove them. 
We are now working to  the improvement of the precision of the 
assembly steps. Also the quality of the mosaic crystals of Cu(111) will be
improved, 
with the goal of building a new Laue lens prototype model in the second 
semester of this year.

\acknowledgments     %>>>> equivalent to \section*{ACKNOWLEDGMENTS}       
 
We acknowledge the financial support by the Italian Space Agency ASI 
and a minor contribution by the Italian Institute of Astrophysics (INAF). 
The design study was also possible 
thanks to the award of the 2002 Descartes Prize of the European Committee.

%%-----------------------------------------------------------
%Citations to the references are made using superscript numerals, as 
%demonstrated in the preceding paragraph.  One may also directly refer 
%to a reference within the text, e.g., ``as shown 
%in Ref.~\citenum{Metropolis53} ..." 

%%%%% References %%%%%

\bibliography{lens_biblio}   %>>>> bibliography data in lens_biblio.bib

\begin{thebibliography}{10}

\bibitem{Ubertini03}
{Ubertini}, P., {Lebrun}, F., {Di Cocco}, G., {Bazzano}, A., {Bird}, A.~J.,
  {Broenstad}, K., {Goldwurm}, A., {La Rosa}, G., {Labanti}, C., {Laurent}, P.,
  {Mirabel}, I.~F., {Quadrini}, E.~M., {Ramsey}, B., {Reglero}, V., {Sabau},
  L., {Sacco}, B., {Staubert}, R., {Vigroux}, L., {Weisskopf}, M.~C., and
  {Zdziarski}, A.~A., ``{IBIS: The Imager on-board INTEGRAL},'' {\em Astron. \&
  Astrophys.}~{\bf 411},  L131--L139 (2003).

\bibitem{Frontera97}
{Frontera}, F., {Costa}, E., {dal Fiume}, D., {Feroci}, M., {Nicastro}, L.,
  {Orlandini}, M., {Palazzi}, E., and {Zavattini}, G., ``{PDS experiment on
  board the BeppoSAX satellite: design and in-flight performance results},'' in
  [{\em EUV, X-Ray, and Gamma-Ray Instrumentation for Astronomy VIII, Oswald H.
  Siegmund; Mark A. Gummin; Eds.}{\nolinebreak\hspace{0.1em}]},  {Siegmund},
  O.~H. and {Gummin}, M.~A., eds., {\em Presented at the Society of
  Photo-Optical Instrumentation Engineers (SPIE) Conference} {\bf 3114},
  206--215 (Oct. 1997).

\bibitem{Knodlseder07}
{Kn{\"o}dlseder}, J., {von Ballmoos}, P., {Frontera}, F., {Bazzano}, A.,
  {Christensen}, F., {Hernanz}, M., and {Wunderer}, C., ``{GRI: focusing on the
  evolving violent universe},'' in [{\em Optics for EUV, X-Ray, and Gamma-Ray
  Astronomy III. Edited by O'Dell, Stephen L.; Pareschi,
  Giovanni.}{\nolinebreak\hspace{0.1em}]},  {\em Presented at the Society of
  Photo-Optical Instrumentation Engineers (SPIE) Conference} {\bf 6688},  5
  (Sept. 2007).

\bibitem{Frontera05a}
{Frontera}, F., {Pisa}, A., {de Chiara}, P., and {et al.}, ``{Exploring the
  hard X-/soft gamma-ray continuum spectra with Laue lenses},'' in [{\em ESA
  Special Publication}{\nolinebreak\hspace{0.1em}]},  {Favata}, F.,
  {Sanz-Forcada}, J., {Gim{\'e}nez}, A., and {Battrick}, B., eds., {\em ESA
  Special Publication} {\bf 588},  323--+ (Dec. 2005).

\bibitem{Frontera06}
{Frontera}, F., {Pisa}, A., {Carassiti}, V., {Evangelisti}, F., {Loffredo}, G.,
  {Pellicciotta}, D., {Andersen}, K.~H., {Courtois}, P., {Amati}, L., {Caroli},
  E., {Franceschini}, T., {Landini}, G., {Silvestri}, S., and {Stephen}, J.~B.,
  ``{Gamma-ray lens development status for a European gamma-ray imager},'' in
  [{\em Space Telescopes and Instrumentation II: Ultraviolet to Gamma
  Ray}{\nolinebreak\hspace{0.1em}]},  {Turner}, M.~J.~L. and {Hasinger}, G.,
  eds., {\em Presented at the Society of Photo-Optical Instrumentation
  Engineers (SPIE) Conference} {\bf 6266} (July 2006).

\bibitem{Knodlseder06}
{Kn{\"o}dlseder}, J., ``{GRI: the gamma-ray imager mission},'' in [{\em Space
  Telescopes and Instrumentation II: Ultraviolet to Gamma Ray. Edited by
  Turner, Martin J. L.; Hasinger, G{\"u}nther. Proceedings of the SPIE, Volume
  6266, pp. 626623 (2006).}{\nolinebreak\hspace{0.1em}]},  {Turner}, M.~J.~L.
  and {Hasinger}, G., eds., {\em Presented at the Society of Photo-Optical
  Instrumentation Engineers (SPIE) Conference} {\bf 6266} (July 2006).

\bibitem{Frontera07}
{Frontera}, F., {Loffredo}, G., {Pisa}, A., {Milani}, L., {Nobili}, F.,
  {Auricchio}, N., {Carassiti}, V., {Evangelisti}, F., {Landi}, L.,
  {Squerzanti}, S., {Andersen}, K.~H., {Courtois}, P., {Amati}, L., {Caroli},
  E., {Landini}, G., {Silvestri}, S., {Stephen}, J.~B., {Poulsen}, J.~M.,
  {Negri}, B., and {Pareschi}, G., ``{Development status of a Laue lens project
  for gamma-ray astronomy},'' in [{\em Optics for EUV, X-Ray, and Gamma-Ray
  Astronomy III. Edited by O'Dell, Stephen L.; Pareschi, Giovanni. Proceedings
  of the SPIE}{\nolinebreak\hspace{0.1em}]},  {\em Presented at the Society of
  Photo-Optical Instrumentation Engineers (SPIE) Conference} {\bf 6688},  20
  (2007).

\bibitem{Pellicciotta06}
{Pellicciotta}, D., {Frontera}, F., {Loffredo}, G., {Pisa}, A., {Andersen}, K.,
  {Courtois}, P., {Hamelin}, B., {Carassiti}, V., {Melchiorri}, M., and
  {Squerzanti}, S., ``{Laue Lens Development for Hard X-rays ($>$60 keV)},''
  {\em IEEE Trans. Nucl. Sci.}~{\bf 53},  253--258 (2006).

\bibitem{Courtois04}
{Courtois}, P., {Hamelin}, B., and {Andersen}, K.~H., ``{Production of copper
  and Heusler alloy Cu$_{2}$MnAl mosaic single crystals for neutron
  monochromators},'' {\em Nuclear Instruments and Methods in Physics Research
  A}~{\bf 529},  157--161 (Aug. 2004).

\bibitem{Abrosimov05}
{Abrosimov}, N.~V., ``{Mosaic and gradient SiGe single crystals for gamma ray
  Laue lenses},'' {\em Experimental Astronomy}~{\bf 20},  185--194 (Dec. 2005).

\bibitem{Barriere08}
{Barriere}, N., {\em {Developpement d'une lentille de Laue pour l'astrophysique
  nucleaire}}, PhD thesis, University of Toulouse, France (2008).

\bibitem{Pisa05b}
{Pisa}, A., {Frontera}, F., {Loffredo}, G., {Pellicciotta}, D., and
  {Auricchio}, N., ``{Optical properties of Laue lenses for hard X-rays ($>$60
  keV)},'' {\em Experimental Astronomy}~{\bf 20},  219--228 (Dec. 2005).

\bibitem{Pisa04}
{Pisa}, A., {Frontera}, F., {De Chiara}, P., {Loffredo}, G., {Pellicciotta},
  D., {Landini}, G., {Franceschini}, T., {Silvestri}, S., {Andersen}, K.,
  {Courtois}, P., and {Hamelin}, B., ``{Feasibility study of a Laue lens for
  hard x rays for space astronomy},'' in [{\em Advances in Computational
  Methods for X-Ray and Neutron Optics. Edited by Sanchez del Rio,
  Manuel.}{\nolinebreak\hspace{0.1em}]},  {Sanchez del Rio}, M., ed., {\em
  Presented at the Society of Photo-Optical Instrumentation Engineers (SPIE)
  Conference} {\bf 5536},  39--48 (Oct. 2004).

\bibitem{Vonballmoos04}
{von Ballmoos}, P., {Halloin}, H., {Skinner}, G.~K., {Smither}, R.~K., {Paul},
  J., {Abrosimov}, N.~V., {Alvarez}, J.~M., {Astier}, P., {Bastie}, P.,
  {Barret}, D., {Bazzano}, A., {Boutonnet}, A., {Brousse}, P., {Cordier}, B.,
  {Courvoisier}, T., {Di Cocco}, G., {Giuliani}, A., {Hamelin}, B., {Hernanz},
  M., {Jean}, P., {Isern}, J., {Kn{\"o}dlseder}, J., {Laurent}, P., {Lebrun},
  F., {Marcowith}, A., {Martinot}, V., {Natalucci}, L., {Olive}, J.-F., {Pain},
  R., {Sadat}, R., {Sainct}, H., {Ubertini}, P., and {Vedrenne}, G., ``{MAX: a
  gamma-ray lens for nuclear astrophysics},'' in [{\em Optics for EUV, X-Ray,
  and Gamma-Ray Astronomy. Edited by Citterio, Oberto; O'Dell, Stephen
  L.}{\nolinebreak\hspace{0.1em}]},  {Citterio}, O. and {O'Dell}, S.~L., eds.,
  {\em Presented at the Society of Photo-Optical Instrumentation Engineers
  (SPIE) Conference} {\bf 5168},  482--491 (Feb. 2004).

\bibitem{Loffredo05}
{Loffredo}, G., {Frontera}, F., {Pellicciotta}, D., {Pisa}, A., {Carassiti},
  V., {Chiozzi}, S., {Evangelisti}, F., {Landi}, L., {Melchiorri}, M., and
  {Squerzanti}, S., ``{The Ferrara hard X-ray facility for testing/calibrating
  hard X-ray focusing telescopes},'' {\em Experimental Astronomy}~{\bf 20},
  413--420 (Dec. 2005).

\end{thebibliography}
\bibliographystyle{spiebib}   %>>>> makes bibtex use spiebib.bst

\end{document}